\documentclass[a4paper,onecolumn,twoside,fleqn,10pt]{article}
\usepackage{amsmath}
\usepackage{authblk}
\usepackage[svgnames]{xcolor}
\definecolor{blue}{RGB}{0,0,255}
\usepackage{caption}
\usepackage[font={color=blue},figurename=Fig.,labelfont={it}]{caption}
\usepackage{bm}
\usepackage{subcaption}
\usepackage[varg]{txfonts}
\usepackage{graphicx}
\usepackage{pdfpages}
\usepackage{wasysym}
\usepackage{rotating}
\usepackage{hyperref}
\usepackage{indentfirst}
\usepackage{float}
\usepackage{lipsum} 
\usepackage{setspace} 
\usepackage{indentfirst}
\usepackage[english]{babel}
\usepackage[switch,columnwise]{lineno}
\usepackage[left=2cm,right=2cm,top=2cm,bottom=2cm]{geometry}

\hypersetup{
    colorlinks=true,                          
    linkcolor=blue, 
    citecolor=blue, 
    urlcolor=blue,
    linktoc=all
}

\newcommand{\blue}[1]{\textcolor{blue}{#1}}

\newcommand{\eg}{\textit{eg }}
\newcommand{\cf}{\textit{cf }}
\newcommand{\ie}{\textit{ie }}

\newcommand{\Poisson}{\blue{Poisson} }

\newcommand{\Lagrange}{\blue{Lagrange} }
\newcommand{\Laplace}{\blue{Laplace} }
\newcommand{\dAlembert}{\blue{d'Alembert} }
\newcommand{\Poincare}{\blue{Poincaré} }

\newcommand{\Fang}{\blue{Fang et al.} }
\newcommand{\Einstein}{\blue{Einstein} }
\newcommand{\Milankovic}{\blue{Milankovic} }
\newcommand{\Pouillet}{\blue{Pouillet} }
\newcommand{\Jose}{\blue{Jose} }
\newcommand{\Gleissberg}{\blue{Gleissberg} }
\newcommand{\Schwabe}{\blue{Schwabe} }
\newcommand{\Hale}{\blue{Hale} }
\newcommand{\Golyandina}{\blue{Golyandina and Zhigljavsky} }
\newcommand{\Barnhart}{\blue{Barnhart and Eichinger} }
\newcommand{\Takalo}{\blue{Takalo and Mursula} }
\newcommand{\Stephenson}{\blue{Stephenson and Morrison} }
\newcommand{\Gross}{\blue{Gross} }
\newcommand{\Keogh}{\blue{Keogh et al. } }
\newcommand{\Lin}{\blue{Lin et al. } }
\newcommand{\Brauning}{\blue{Br{\"a}uning } }
\newcommand{\Leroy}{\blue{Le Roy Ladurie and Vasak  } }

\usepackage{fancyhdr}
\begin{document}

\title{A living forest of Tibetan Juniper trees as  a new kind of astro-geophysical observatory}

% \Author[affil]{given_name}{surname}

\author[1]{Courtillot Vincent}
\author[2]{Boulé Jean-Baptiste}
\author[1]{Le Mouël Jean-Louis}
\author[3]{Gibert Dominique}
\author[4]{Zuddas Pierpaolo}
\author[4]{Maineult Alexis}
\author[5]{Gèze Marc}
\author[1]{Lopes Fernando}

\affil[1]{Université Paris Cité, Institut de Physique du globe de Paris, CNRS UMR 7154, F-75005 Paris,France}
\affil[2]{CNRS UMR7196, INSERM U1154, Museum National d’Histoire Naturelle, Paris, F-75005, France}
\affil[3]{LGL-TPE - Laboratoire de Géologie de Lyon - Terre, Planètes, Environnement, Lyon, France}
\affil[4]{Sorbonne Université, CNRS, METIS, F75005, Paris, France}
\affil[5]{Muséum National d’Histoire Naturelle, Sorbonne Université, CEMIM, Paris, F-75005, France}

\date{\today}
\maketitle

\begin{abstract}
The elliptical trajectory of Earth about the Sun is perturbed by torques exerted by the Moon and Sun, and also the four giant planets. These provoke variations of insolation at Earth surface, known as kyr-long Milanković cycles. The concept has been extended to the shorter time scales of years to centuries, that are relevant to tree growth. This paper focuses on iterative Singular Spectrum Analysis (iSSA) of results of the dendro-chronological study of a forest of long-lived Tibetan junipers, made available by \Fang \cite{fang2018}. From this, we determine a median curve of tree growth rates TRGRm, that is analyzed by iSSA. We obtain a rich set of (pseudo-) periods, from 3.3 yr up to more than 1000 years, that compare favorably with the specific spectral signature (SSS) found in the sunspot and length-of-day time series. We discuss in detail the record from a single tree that spans almost completely the 357-2000 AD interval. The 90 yr Gleissberg, 22 yr and 30 yr components are quite prominent. The Oort, Wolf, Spörer, Maunder and Dalton climate extrema all correspond quite precisely to extrema of the Gleissberg cycle. The well-known Medieval Climate Optimum (MCO), Little Ice Age (LIA) and Modern Climate Optimum all seem to be mainly forced by variations in the envelope of the Gleissberg cycle. The Gleissberg cycle is strongly modulated with a period of ~500-600 years. The node near a small gap in the data is very close to the Medieval Climate Optimum. Observations in different parts of Earth are in favor of a global extension of the MCO. In the same way that the Milankovic mathematical theory of climate allows one to relate climate change and length of day, through changes in inclination of Earth’s rotation axis and solar insolation, it is reasonable to propose that the set of pseudo-periods that are evidenced in the Tibetan tree ring growth rates simply corresponds to short period Milankovic cycles. The Dulan forest could be considered as a good candidate for a continuous, global geophysical observatory that could monitor polar motion and variations in insolation.
\end{abstract}

\Einstein \cite{einstein1912} was the first to propose a physical explanation for photosynthesis, the main photochemical process through which plants (including trees) develop. Einstein wrote in his introduction "that the number of molecules decomposed per unit time is proportional to the intensity of the active radiation", thus explaining (among many other consequences) the seasonal alternation of the width of tree rings. The elliptical trajectory of our planet about the Sun is perturbed by torques exerted primarily by the Moon and the Sun, and also by the four giant (Jovian) planets. These perturbations provoke slow variations of insolation at the Earth surface. This astronomical phenomenon, known as the \Milankovic \cite{milankovic1920} mathematical theory of climate, is responsible for climate variations (including ice ages) on time scales of thousands of years and much more (\eg \cite{laskar2004,lopes2021a,walzer2023}). One can find a rather simple equation on page 15 of \Milankovic \cite{milankovic1920} thesis that links the amount of insolation ($\mathcal{W}$), the distance from the Sun ($\rho$), its declination ($\delta$)and the so called "solar constant" ($\mathcal{I}_0$), for any geographical location with longitude ($\psi$) and latitude ($\varphi$),
\begin{equation}
	\dfrac{d \mathcal{W}}{dt} = \dfrac{\mathcal{I}_0}{\rho ^2} [\textrm{sin}\ \varphi \ \textrm{sin}\ \delta + \textrm{cos}\ \varphi \ \textrm{cos}\ \delta \ \textrm{cos}\ (\omega + \psi) ].
	\label{eq:01}
\end{equation}

	We have recently extended the concept of \Milankovic \cite{milankovic1920} cycles to the much shorter time scale of years to centuries (\cf \cite{lopes2022a}). The current estimate of the solar ‘constant’ is approximately 1368 $W.m^{-2}$. Since the first pioneering measurement by \Pouillet \cite{pouillet1837} at 1230  $W.m^{-2}$, $\mathcal{I}_0$ has fluctuated by no more than a few percent (\eg \cite{abbot1911,johnson1954,gueymard2018}). This small and slow radiation stands in contrast with the solar activity as monitored by the sunspot cycles of $\sim$11 years \cite{hathaway2015,usoskin2017}. 
	
	A remarkable dendro-chronological study of a forest of Tibetan junipers, recently made available by \Fang \cite{fang2018}, provides a unique opportunity to refine our quantified understanding of a number of terrestrial and solar influences on plant growth, and in return to ascertain the influence of angular momentum exchanges between the planets. We start by recalling the existence of a specific set of spectral (SSS) components that affect many solar and terrestrial phenomena (section \ref{sec:02}). In section \ref{sec:03}, we appeal to Lagrangian mechanics to explain the orbital perturbations that affect all planets axes of rotation. A summary of the Tibetan data acquired by \cite{fang2018} follows (section \ref{sec:04}), from which we determine a median curve of tree growth rates. This time series in then analyzed (section \ref{sec:05}) using the iterative version of Singular Spectrum analysis (section \ref{sec:06}), which is tested against the method of continuous wavelet transforms. We obtain a set of (pseudo-) periods (section \ref{sec:07}). We compare that list to the specific spectral signature SSS found in the sunspot (section \ref{sec:08}) and length of day (section \ref{sec:09}) time series. After a short section \ref{sec:10} on the 11-yr Schwabe cycle as recorded in the tree ring median curve, we discuss in more detail the remarkable record from a tree that spans almost completely the 357 to 2000 AD interval of time (\ref{sec:11}). In the last section \ref{sec:12}, devoted to further discussion and to our conclusions, we discuss how such a forest could be considered as a continuous, global geophysical observatory that could monitor polar motion and variations in insolation.

\section{A specific spectral signature for sunspots and Earth rotation: \label{sec:02}}
	Actually, 70\% of the variance (energy) carried by the sunspots can be broken down (\eg \cite{usoskin2017,lemouel2020,courtillot2021}) into the sum of a nonlinear "trend" (that could be part of a longer \Jose \cite{jose1965} cycle of 165 years), a $\sim$90 yr \Gleissberg \cite{gleissberg1939} cycle (\eg \cite{lemouel2017}), and the famous $\sim$11 yr (actually 9 to 15 yr) \Schwabe \cite{schwabe1844} quasi-cycle. These three components have a similar intensity.
	
	Equation (\ref{eq:01}) can be read in the two directions: from right to left one sees that the inclination of the Earth's axis of rotation forces the insolation received at any location and time scale. It has been shown in a series of papers (\eg \cite{lopes2017,lemouel2019a,lopes2021b,lopes2022b}) that complete (3D) polar motion (that is the sum of (2D) motion of the rotation pole and (1D) length of the day) could be broken down into the sum of pseudo-cycles that are all also found in the sunspot series (\eg \cite{sugawa1972,djurovic1996,del2003,ma2007,chapanov2017,lemouel2020,courtillot2021}); and these correspond more or less (rather more than less) to the periods of revolution of the planets (mainly the giant ones) and to their commensurable periods (\eg \cite{morth1979,morner2013,bank2022,scafetta2023}). We have been able to generalize these "correlations" (that is the recognition of a specific spectral signature SSS in many geo- and astro-time series), that are undoubtedly observations, to a large number of other observed geophysical time series; some had been known in the scientific literature for a long time (\eg \cite{lyons1899,malburet1919,bartels1932,morth1979}). But one can now propose an explanation for these correlations (or rather common, specific spectral signatures).
	
\section{Appealing to Mechanics from the Enlightenment \label{sec:03}}
	\dAlembert \cite{dalembert1749}, \Lagrange \cite{lagrange1788}, \Laplace \cite{laplace1799}, \Poisson \cite{poisson1826} and \Poincare \cite{poincare1893} all contributed to the construction of a physical theory applicable to general geophysics, and particularly the influence of the angular momenta of the planets in our solar system on Earth. These contributions that are too often underestimated today, they have recently studied in some detail (\eg \cite{scafetta2014,scafetta2016,stefani2019,dumont2020,zaccagnino2020,lopes2021b,dumont2022a,dumont2022b,horstmann2023,lemouel2023a,lopes2023a,nesi2023,scafetta2023}). In short, the planets that revolve about our Sun perturb each other's orbits through an exchange of angular momentum. When one solves the Lagrangian of planetary motion (\cite{lagrange1788,lopes2023a}), this second order perturbation of the gravity field results in a delay of perihelion whose mathematical formula is asymptotically identical to that obtained by \Einstein \cite{einstein1915}. We have shown that there is a linear relation between delay in perihelion and the ratio of rotation to revolution periods. The Lagrangian interpretation is rather simple, similar to the weight of a top acting on its rotation axis \cite{lagrange1788}. All planetary torques act (to second order) on the axes of rotation of all planets in general, and that of Earth in particular: the Earth'inclination is perturbed, hence its rotation and revolution \cite{laplace1799}. 
	
\section{Solar Activity and/or Polar Motion ? The case of Tibetan Junipers\label{sec:04}}
	The two mechanisms summarized above, solar activity and polar motion, could explain all or part of variations in amount of insolation (photons) reaching our planet ($\mathcal{W}$) at any geographical location ($\psi$,$\varphi$). They must therefore influence plant and tree growth. It has been known for a long time that past solar activity can be monitored by performing radiogenic analysis of tree rings (\eg \cite{eddy1976,suess1980,usoskin2017,brehm2021,miyake2021}). A remarkable dendro-chronological study of a forest of Tibetan junipers, recently made available by \Fang \cite{fang2018}. is at the focus of the present paper.
	
	 These authors have sampled 469 junipers (\textit{Juniperus przewalskii Kom} and \textit{Juniperus tibetica Kom}) from 11 sites located on the northern
Tibetan plateau, at an elevation ranging from 3000 to 4500 meters (see \cite{fang2018}, figure 01). The remote altitudinal location ensures minimal anthropologic influence. The tree ages span from 56 to 2008 AD. Indeed, these trees are known to live up to thousands of years (\eg \cite{piovesan2021}). 499 samples have been analyzed to a resolution of 1 micron and dated at the Rinntech laboratory in Heidelberg (Germany). The 499 curves of growth rate of tree rings (in mm/yr) are shown in Figure \ref{fig:01}. 
	 
	 	We first build from the data of \Fang \cite{fang2018} a median curve of growth rates that in a sense best represents the data of the ensemble of Tibetan junipers (in that case data from the last 400 years are used). We next decompose that curve in a suite of components, using iterative Singular Spectrum Analysis (iSSA) in order to detect and extract a trend and quasi-periodic components. After discussing aspects of the method, we compare the list of quasi-cycles in the median Tibetan curve with its equivalent in the motion of the complete pole of rotation and that for the global mean temperature.
	 	
\section{Time Series Analysis of the Tibetan Junipers \label{sec:05}}
\begin{figure}[H]
		\centering{\includegraphics[width=0.8\columnwidth]{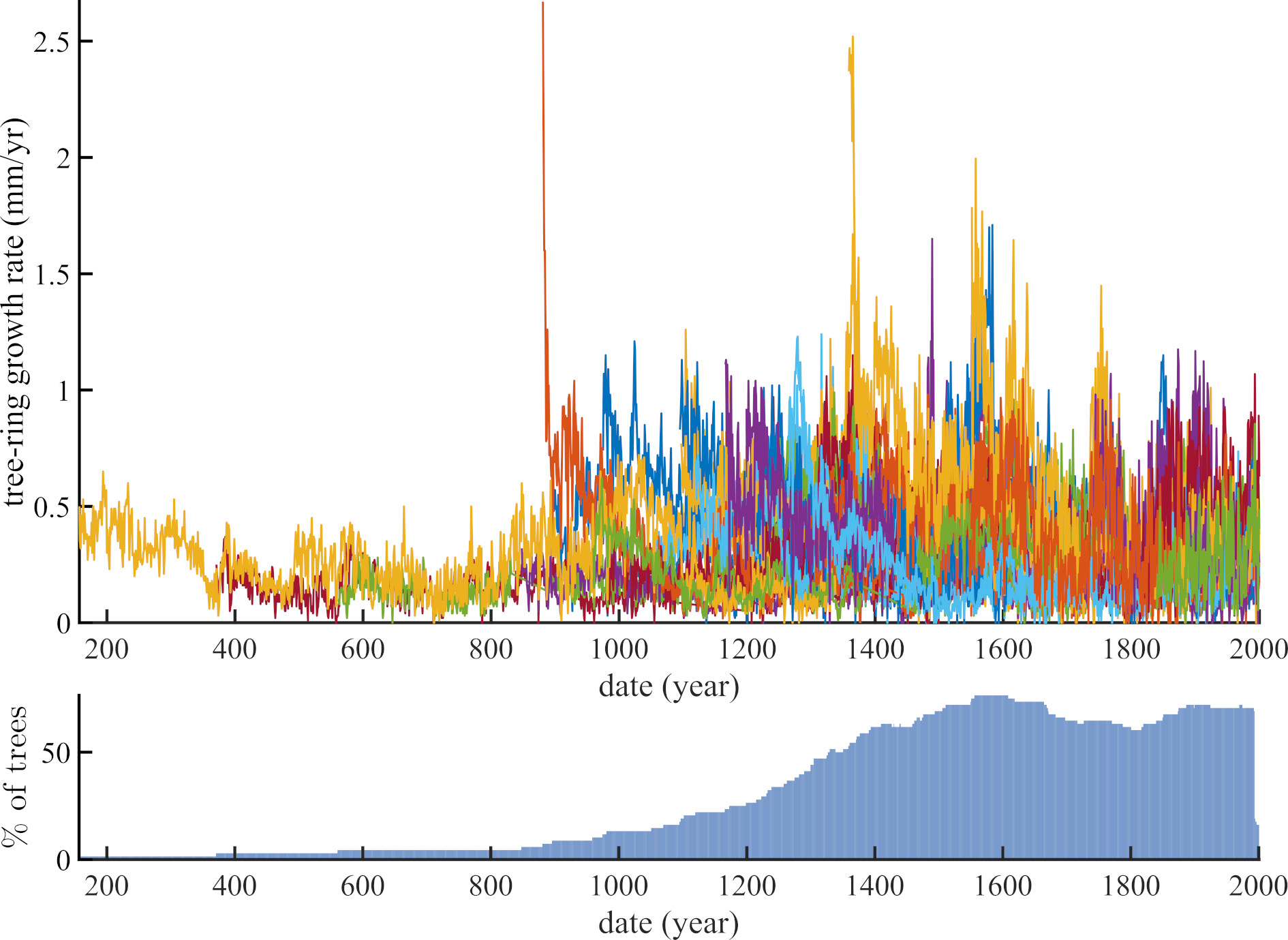}} 	
     	\caption{(top) All (data) curves of growth rates of Tibetan juniper tree-rings (from Fang et al. 2018) (bottom) Histogram of the number of trees as a function of age.}
		\label{fig:01} 	
\end{figure}

\begin{figure}[H]
		\centering{\includegraphics[width=0.8\columnwidth]{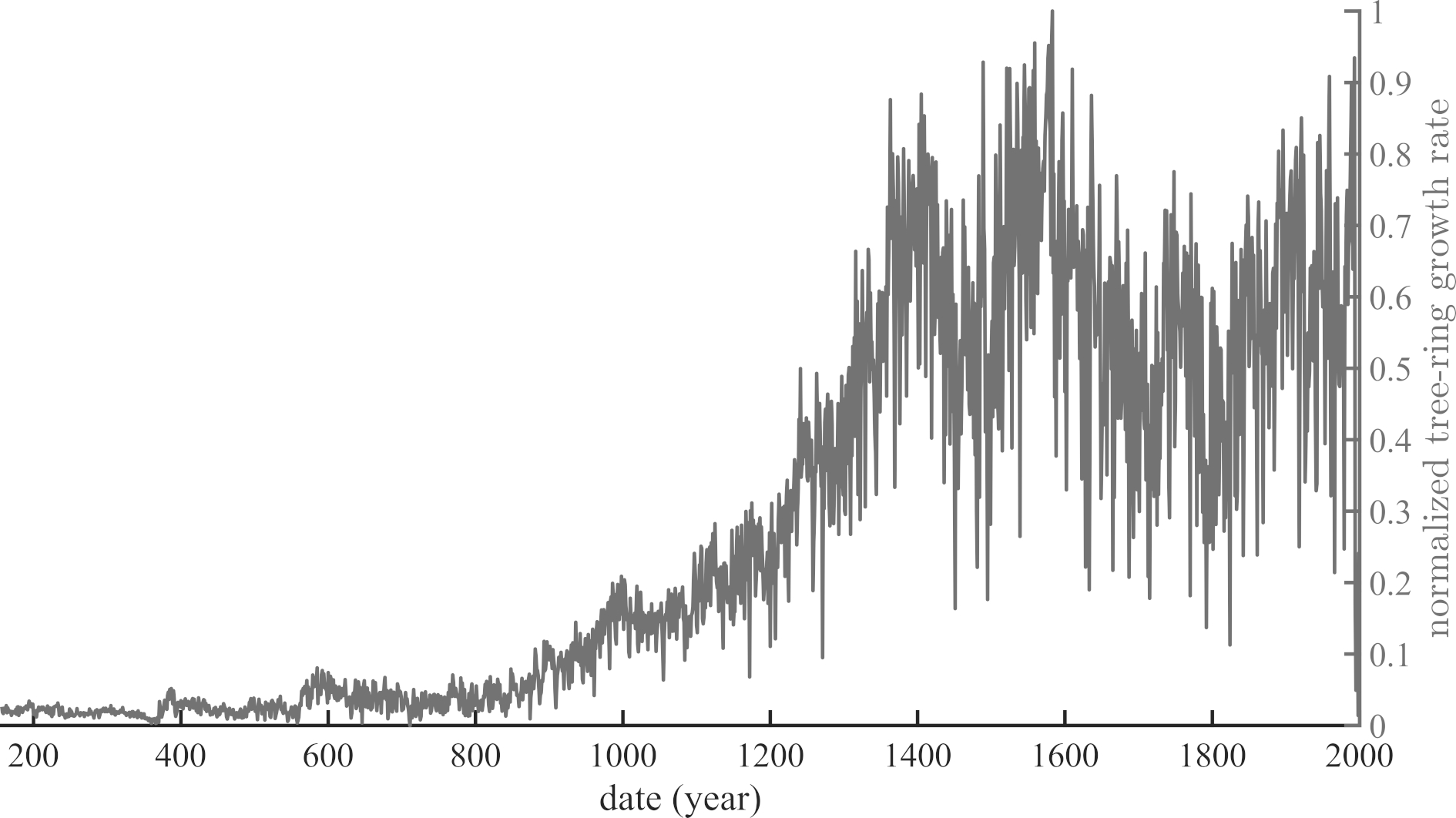}} 	
     	\caption{Median curve constructed from the data shown in Figure \ref{fig:01}}
		\label{fig:02} 	
\end{figure}

	The top part of Figure \ref{fig:01} displays the raw data from Dulan site; the lower part displays the histogram of the number of tree samples of a given age. Because the number of trees available evolves with age, the final median curve of tree-ring growth rate has been normalized: the value of 100\% corresponds to the total number of individuals for the entire period covered by samples, that is 469 individuals between 57 and 2008 AD. As in previous studies of the world tide gauges \cite{courtillot2022} and of all volcanic eruptions since 1700 \cite{lemouel2023b}, we have decomposed the data in 1952 annual bins, and for each one computed a median growth rate (cf. Figure \ref{fig:02}). The trend of our median curve TRGRm is directly linked to the number of individual samples and is not shown here nor used in the following. As shown in the histogram (Figure \ref{fig:01}, lower part), and in TRGRm (Figure  \ref{fig:01}), the number of data remains stable in the 1400 to 2000 time range. In the next section we focus on these six centuries.

\section{Singular Spectrum Analysis\label{sec:06}}
	The method of spectral analysis Singular Spectrum Analysis (\cf \cite{vautard1989,vautard1992}), comes from geophysical studies as do most methods of time series analysis. It has been developed in order to answer a number of questions that Fourier or wavelet analyses could not answer as quickly or straightforwardly as desired. However, none of these methods is superior to the others in all cases. Given sufficient work they all lead to the same final answer.
	
	The SSA method is fully covered in the reference book by \Golyandina \cite{golyandina2013}. It uses the properties of information embedding of the Hankel and Toeplitz matrices (\eg \cite{lemmerling2001}), that are diagonally descending matrices. These matrices are analyzed through the prism of Singular Value Decomposition (SVD, \cf \cite{golub1971}): the starting time series (signal) is decomposed in a trend (that may not exist) and a sum of pseudo-periodical oscillations. The path is from data space to data space.
	
\begin{figure}[H]
		\centering{\includegraphics[width=0.8\columnwidth]{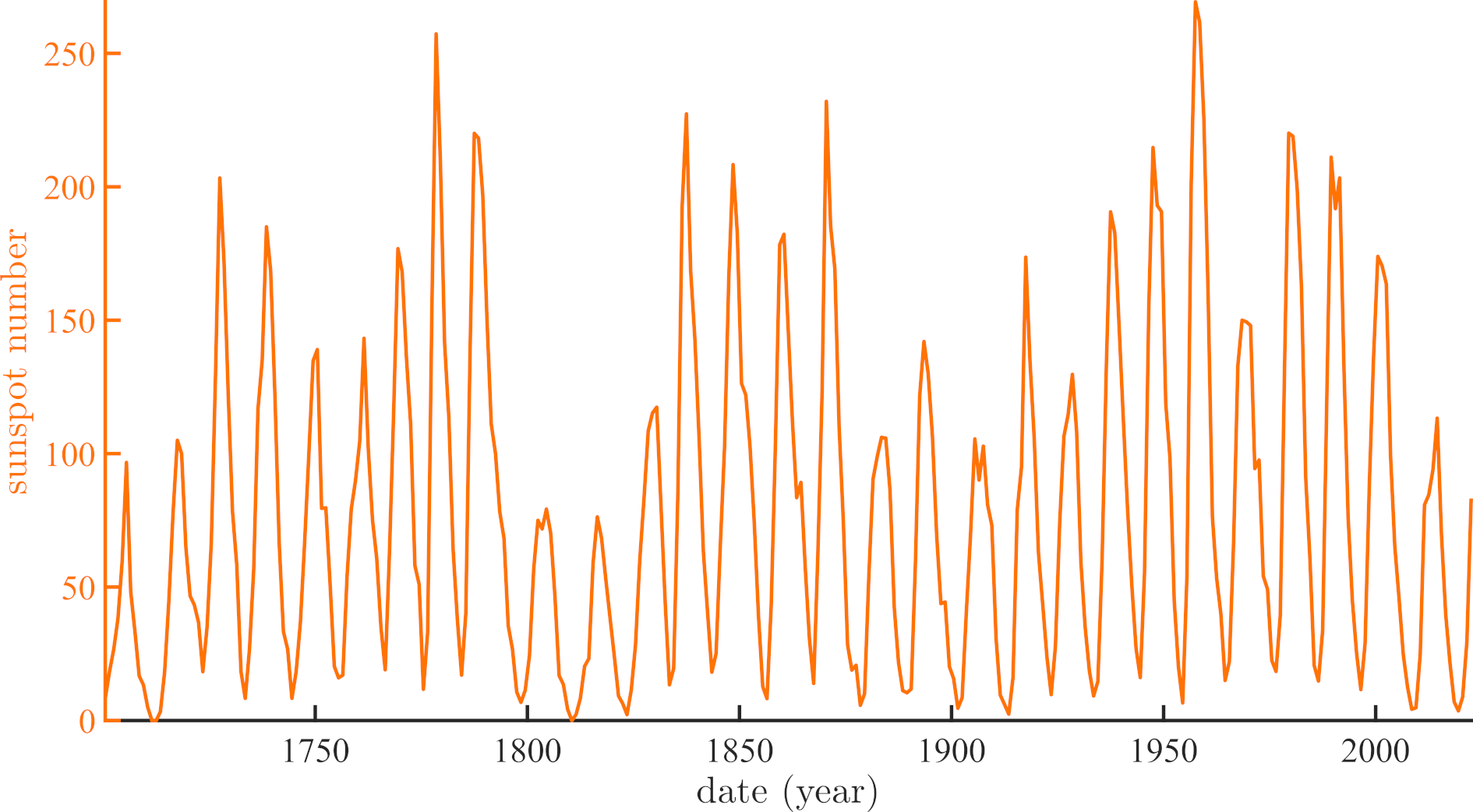}} 	
     	\caption{Annual sunspot numbers since 1700 from the SILSO database.}
		\label{fig:03} 	
\end{figure}

	For illustrative reasons, we start by applying SSA to the series of annual values of sunspot numbers, available from 1700 onwards (other similar case studies of sunspot numbers are found in \Barnhart \cite{barnhart2011} or in \Takalo \cite{takalo2018}, and give the same results as found in the present paper). The data are maintained and distributed by the Royal Observatory of Belgium (SILSO); they are shown in Figure \ref{fig:03}.	
	
	In order to test the quality and validity of the SSA decomposition, we compare its application to sunspot numbers using the continuous wavelet transform \cite{gibert1998}.  
\begin{figure}[H]
		\centering{\includegraphics[width=0.8\columnwidth]{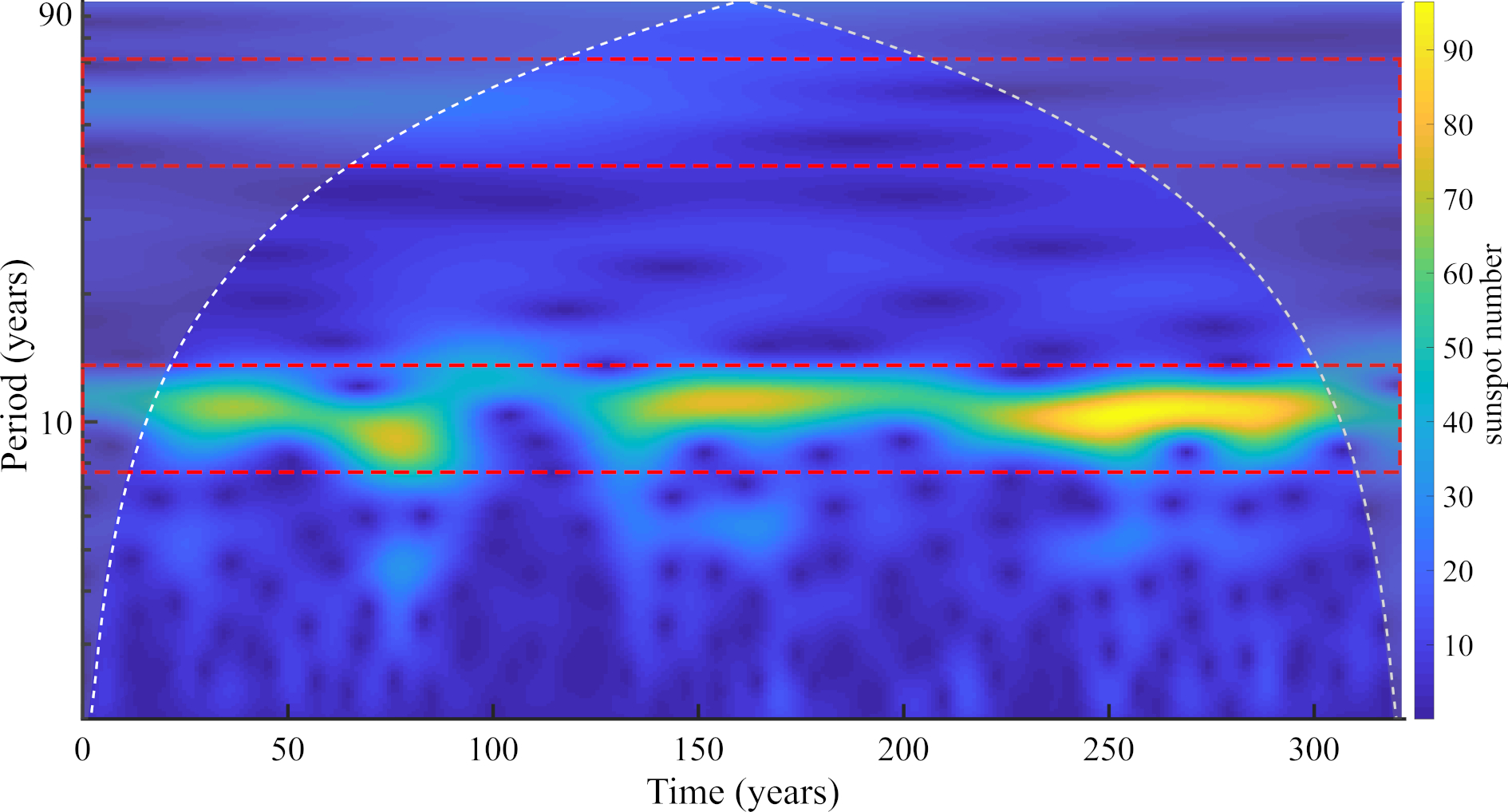}} 	
     	\caption{Scalogram of the sunspot series.}
		\label{fig:04} 	
\end{figure}		

\begin{figure}[H]
		\centering{\includegraphics[width=0.8\columnwidth]{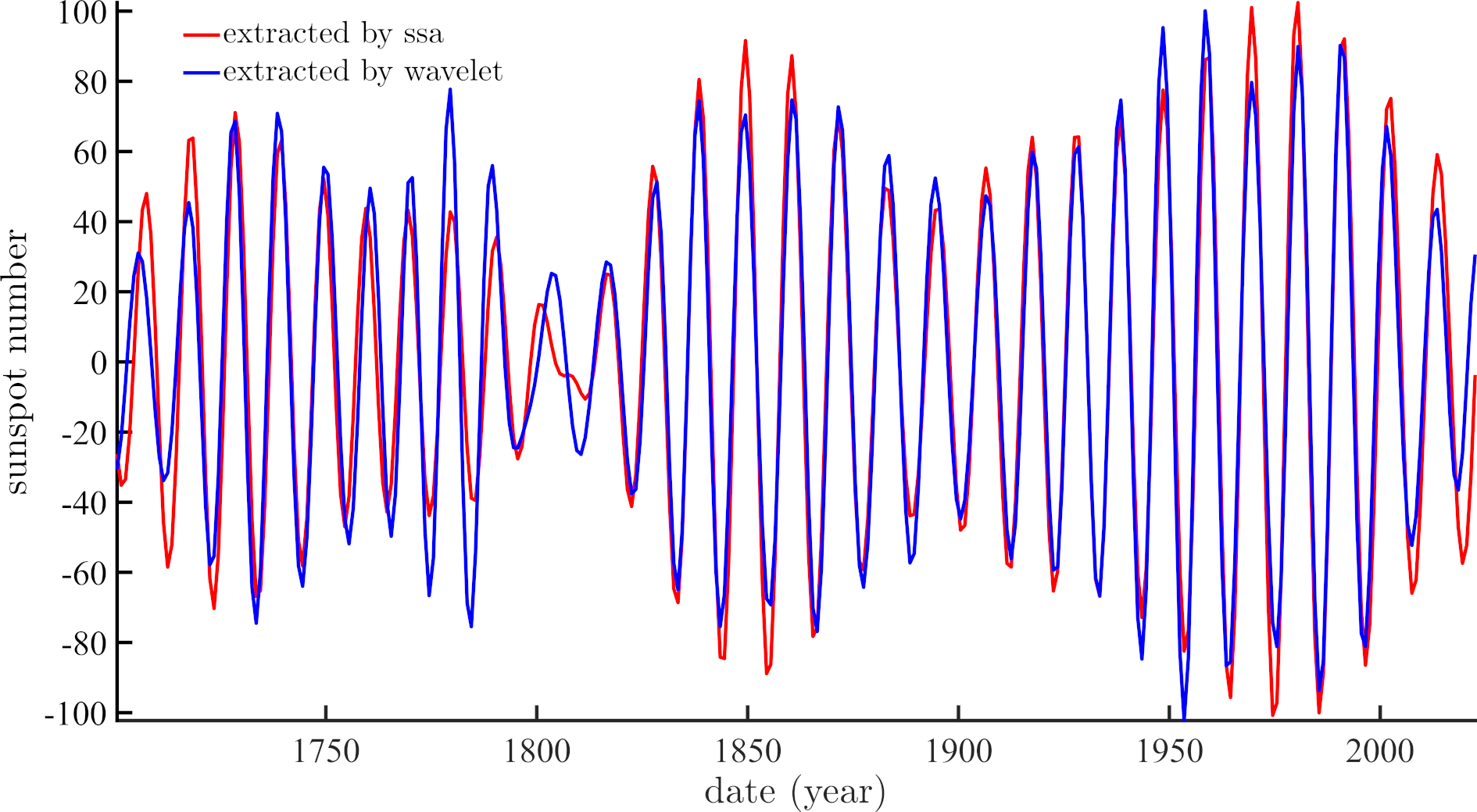}} 	
     	\caption{The Schwabe ($\sim$11 yr) component of the sunspot series extracted using the iSSA (red curve) and the wavelet (blue curve) methods.}
		\label{fig:05} 	
\end{figure}

	Figure \ref{fig:04} shows the scalogram, \ie the continuous wavelet transform of the sunspot series SSN since 1700 (using here a Morlet wavelet). The wavelet coefficients (the ordinates) have been converted to years. The color code indicates the number of sunspots (right scale). The band with long yellow patches corresponds to the famous $\sim$11-yr \Schwabe \cite{schwabe1844} cycle. The component periods actually spread between 9 and 15 years (dashed red horizontal lines). We have also marked in dashed red lines the less conspicuous $\sim$60-yr component. The dashed white curve that looks like an upside down boat marks the boundary where significant spurious edge effects may arise. The wavelet components are extracted following the crests of the bands, using the redundancy property of wavelet information kernels to reconstruct the corresponding signal. This has been done for the Schwabe component (blue curve in Figure \ref{fig:05}). The corresponding component extracted by iSSA is shown in the same Figure (red curve). The match is excellent (except for small edge-effects in the first and last decade of the series) and confirms the quality of both spectral analysis methods.
	
\begin{figure}[H]
		\centering{\includegraphics[width=0.8\columnwidth]{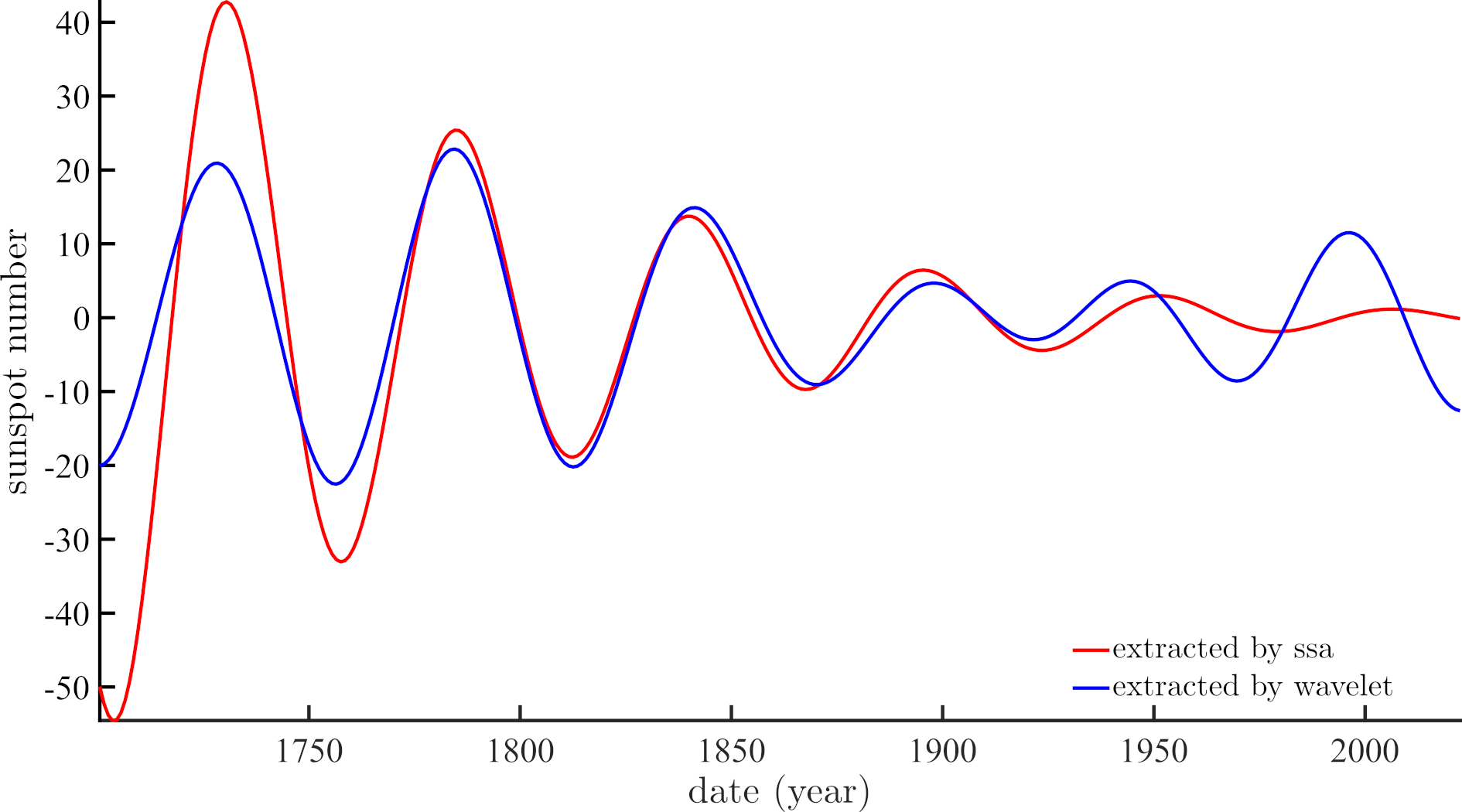}} 	
     	\caption{Same as Figure \ref{fig:05} for the 60-yr cycle}
		\label{fig:06} 	
\end{figure}

		Figure \ref{fig:06} shows the same treatment being applied to the $\sim$60-yr component. As expected for a 60-yr period extracted from a data window of only 300 years, the edge effects are more serious. Still, from 1730 to 2000 the components computed by the two methods agree very well.
 	
	Once all the pseudo-cycles are detected and extracted, we evaluate the corresponding period using Fourier transform, along with the associated error (half-width at half-maximum of the spectrum).
 	
\section{Extracting the set of SSS pseudo-cycles from TRGRm \label{sec:07}}
	Applying iSSA to the series of Tibetan juniper tree ring growth rates generates a set of pseudo- (or quasi-) cycles that are listed in Table \ref{tab:01}. The second column lists our estimates of the uncertainty that should be assigned to the value of the central period of each line (component). The third column indicates whether the same period is found in the sunspot series. Indeed, ten components are common to the two series: they are the \Jose \cite{jose1965}, \Gleissberg \cite{gleissberg1939}, \Hale \cite{usoskin2017} , \Schwabe \cite{schwabe1844} cycles and some of their harmonics. In order to make the paper lighter, we defer a description of the tree ring pseudo-cycles to the Appendix \ref{sec:A}. Recall that these components are determined for the six  centuries since 1400 AD.
	
\begin{table}[H]	
\begin{tabular}{p{4cm}|p{4cm}|p{4cm}}
 \multicolumn{3}{c}{} \\
 \hline
  \hline
 List of extracted pseudo-cycles (year) & errors (year)  & Solar Cycle name \\
 \hline
 \ & \ \\
1006 & undef & \\
527.32 & $\pm$ 109.38 & \\
369.86 & $\pm$ 50.35& \\
232.37 & $\pm$ 60.13 & \\
195.94 & $\pm$ 31.84 & \\
158.84 & $\pm$ 15.18 &  Jose (1965) \\
130.11 & $\pm$ 11.28 & \\
85.04 & $\pm$ 8.48 & Gleissberg (1939)\\
53.03 & $\pm$ 3.02 &  Usoskin (2017)\\
33.55 & $\pm$ 2.27 & Usoskin (2017)\\
21.43 & $\pm$ 0.36 & Hale, Usoskin (2017)\\
15.54 & $\pm$ 0.24 & Schwabe (1844)\\
11.48 & $\pm$ 0.15 & Schwabe (1844)\\
12.70 & $\pm$ 0.13 & Schwabe (1844)\\
8.45 & $\pm$ 0.07 & Schwabe (1844)\\
7.73 & $\pm$ 0.05 & \\
6.76 & $\pm$ 0.04 & \\
3.38 & $\pm$ 0.01 & Schwabe (1844) harmonic\\
 \hline
 \\
\end{tabular}   
	\caption{Column 1: list of pseudo-periods extracted by iSSA from series TRGRm. Column 2: their uncertainties (component width at mid-height). Column 3: component present in the sunspot series.}
	\label{tab:01}
\end{table}	
	
\section{Comparison with the sunspot series SSS \label{sec:08}}
\begin{figure}[H]
		\centering{\includegraphics[width=0.8\columnwidth]{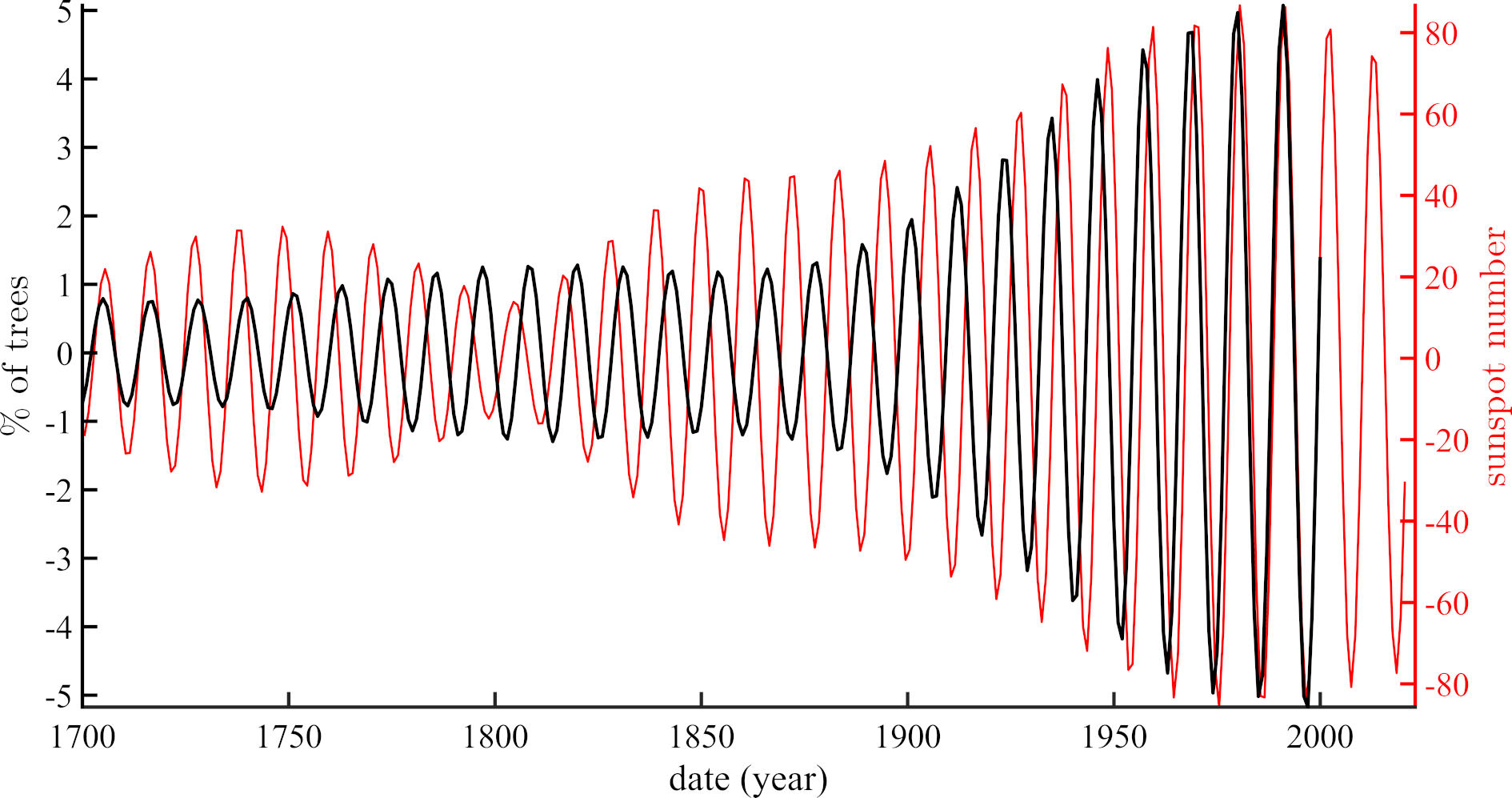}} 	
     	\caption{Superimposition of Schwabe cycles extracted by iSSA from the sunspot series (red curve) and Tibetan tree rings (growth rate) TRGRm (black curve), from 1700 to 2023.}
		\label{fig:07} 	
\end{figure}

	As noted above, the three main components of the sunspot series are the trend and the \Schwabe \cite{schwabe1844} and \Gleissberg \cite{gleissberg1939} cycles. These are also prominent in the tree ring series. Figure \ref{fig:07} shows the superimposition from 1700 to 2000AD of the $\sim$11-yr cycles extracted from sunspots (red curve) and tree rings TRGRm (black curve). The amplitudes of the envelopes of these two Schwabe series evolve in parallel. The two components are in phase from 1700 to 1740, then drift to reach phase opposition around 1890 and are back in conjunction in 1980. This modulation suggests a $\sim$400-yr pseudo-period. 
\begin{figure}[H]
		\centering{\includegraphics[width=0.8\columnwidth]{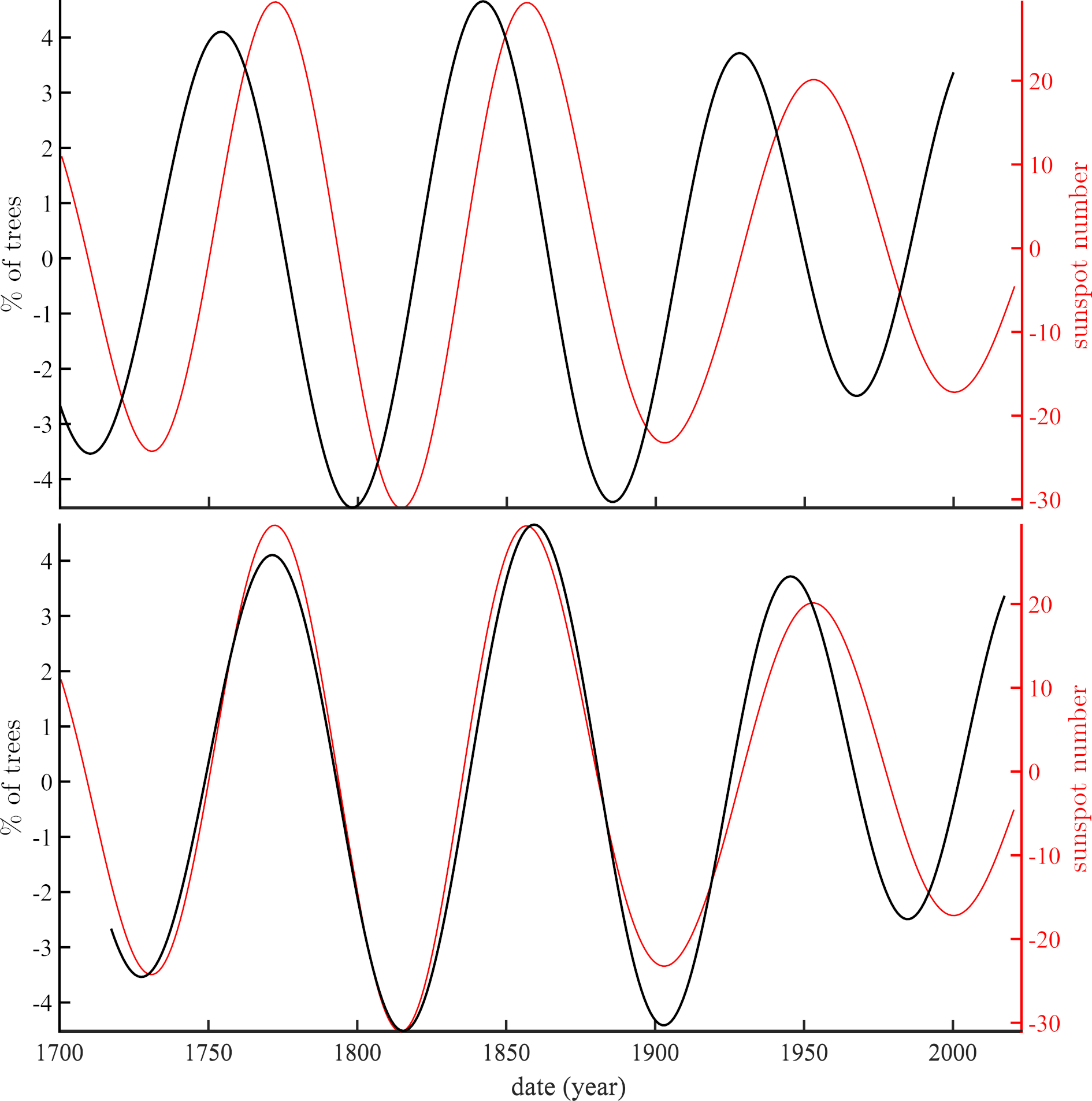}} 	
     	\caption{(Top), same as Figure {fig:07} for the 90-yr cycle. (Bottom), same as with the TRGRm component offset by 12 years.}
		\label{fig:08} 	
\end{figure}	
	
\section{Comparison with the Length of the Day \label{sec:09}}
\begin{figure}[H]
		\centering{\includegraphics[width=0.8\columnwidth]{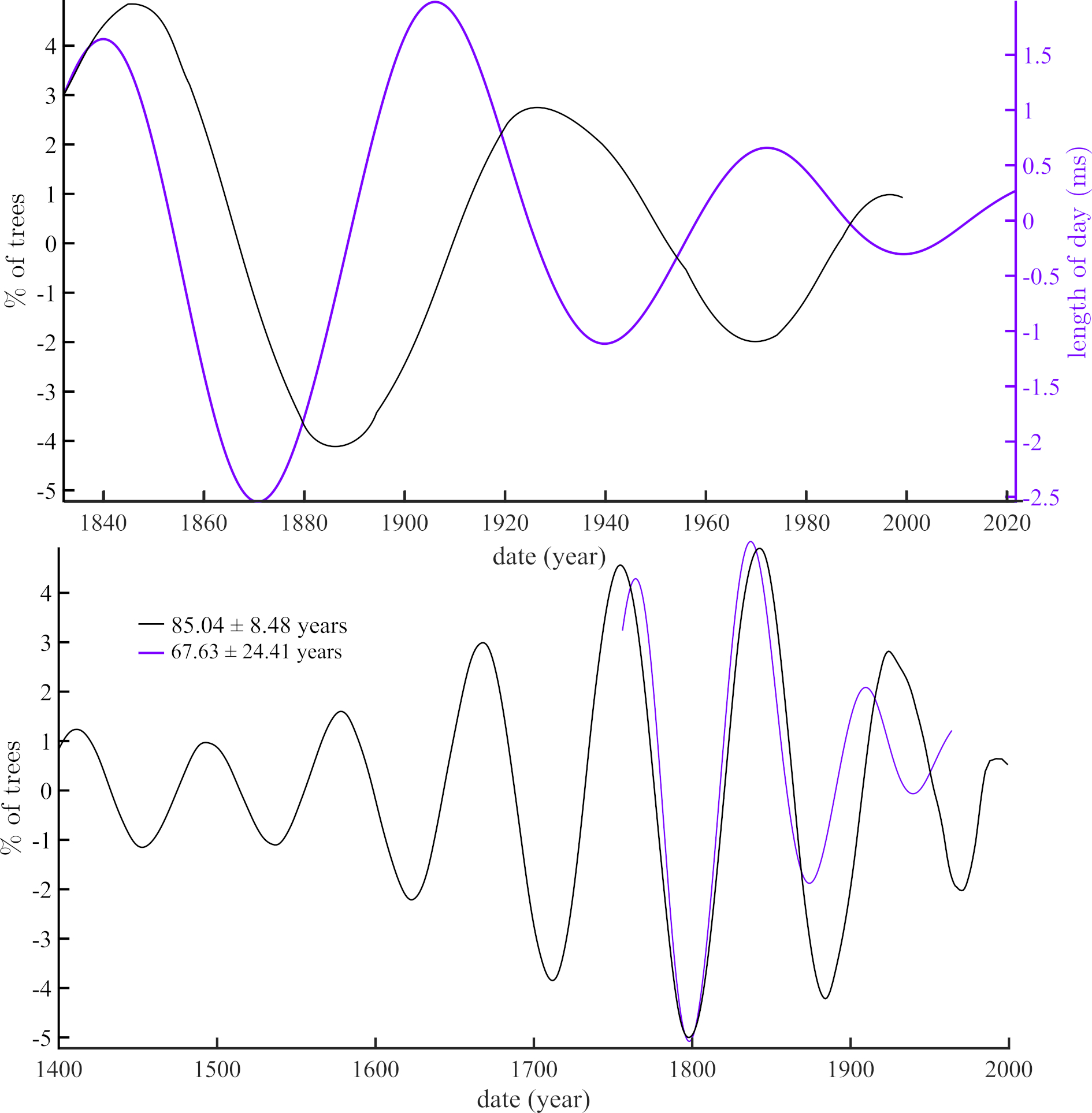}} 	
     	\caption{(Top) superimposition of the Gleissberg $\sim$90 yr component extracted by iSSA from the Gross (2001) length of the day (purple curve) and the Tibetan tree TRGRm (black curve), between 1832 and 2023. (Bottom) Same for the past 6 centuries, from 1400 to 2000. The lod curve has been placed on the time axis to better visualize the phase coherency of the two time series.}
		\label{fig:09} 	
\end{figure}

\begin{figure}[H]
		\centering{\includegraphics[width=0.8\columnwidth]{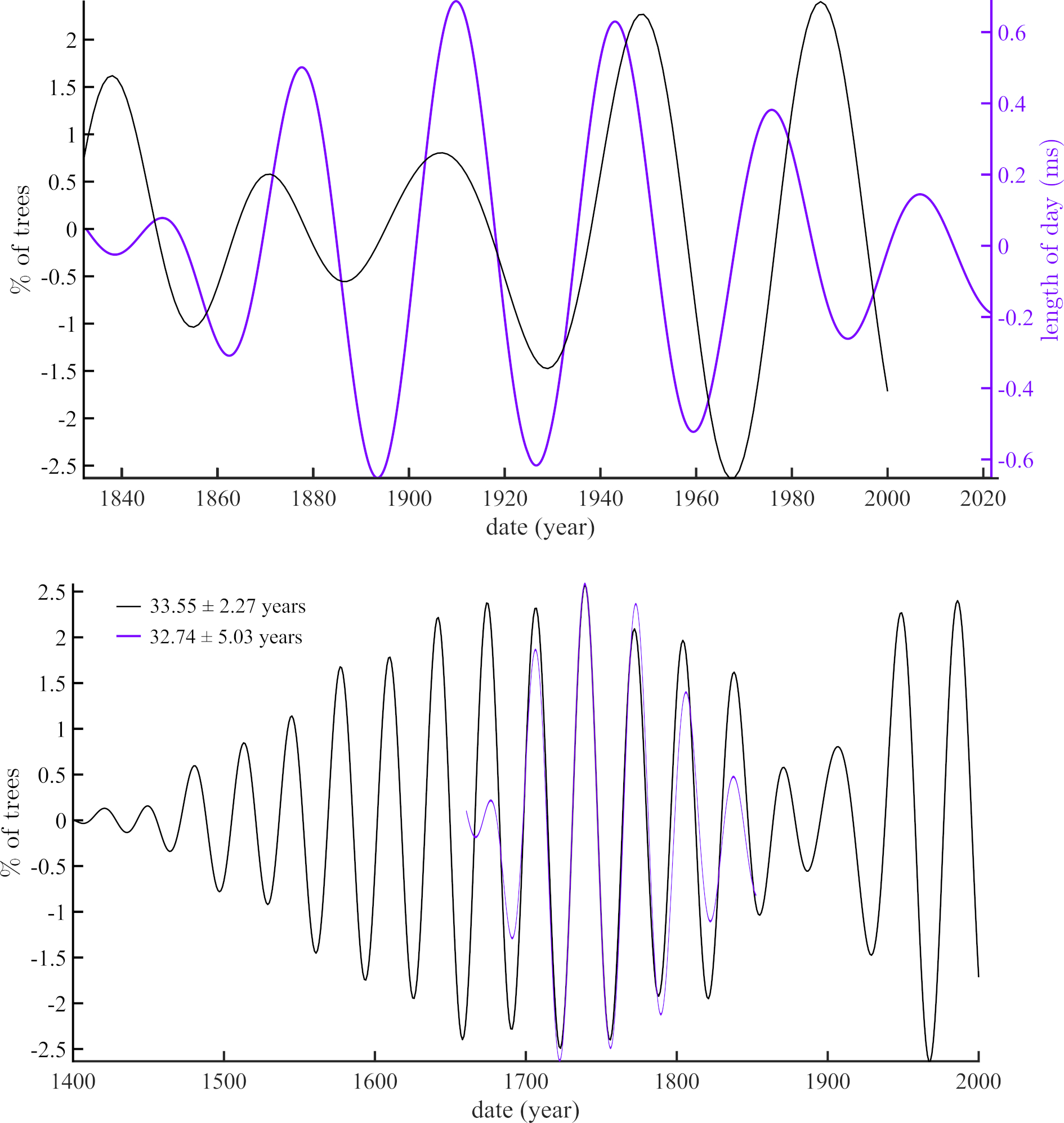}} 	
     	\caption{Same as Figure 11 for the 30-yr cycle.}
		\label{fig:10} 	
\end{figure}

\begin{figure}[H]
		\centering{\includegraphics[width=0.8\columnwidth]{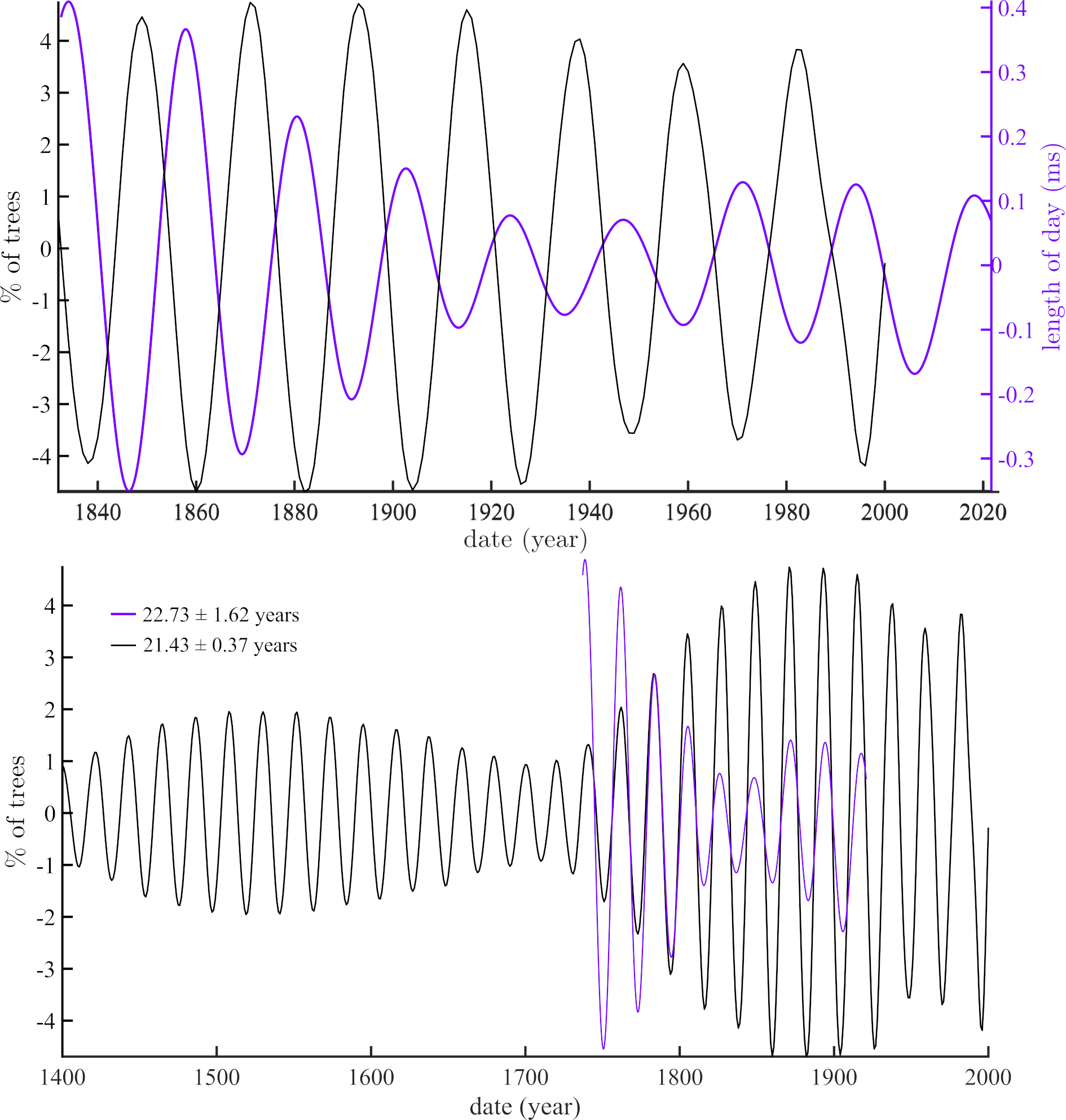}} 	
     	\caption{Same as Figure 09 for the Hale (22 yr) cycle.}
		\label{fig:11} 	
\end{figure}

	\Laplace \cite{laplace1799} was the first to show that the length of the day is identical to the derivative of rotation pole motion (\eg  \cite{lopes2022b,lemouel2023a,lemouel2023a}). We already studied the lod data, namely the compilation of observations of \Stephenson \cite{stephenson1984}, as complemented by \Gross \cite{gross2001} between 1832 and 1997 and by the IERS series of the rotation pole provided daily by satellites since 1962 (\eg  \cite{lopes2022b,courtillot2023}). We present the results of the iSSA analysis of these rotation data for three components, the Gleissberg (Figure \ref{fig:09}), 30 year (Figure \ref{fig:10}) and Hale (Figure \ref{fig:11}) cycles. In each of these three Figures, the top part shows the superimposition of the lod and Tibetan tree TGRGm from 1832 to 2023 and the bottom part shows the full 600 years  from 1400 to 2000, when the number of Juniper samples remains rather constant. In this latter case, the lod component (purple curve) is placed arbitrarily on the time axis, so that the oscillations of lod be in phase with TGRGm. For the three components with periods of 90, 30 and 22 years, lod and TGRGm are in phase and are modulated in similar patterns. In order to save space we do not display the other components but they all agree in the same way.
	
	The comparisons of curves in Figure \ref{fig:09}, \ref{fig:10} and \ref{fig:11} are intended to estimate the pseudo-periods and uncertainties for the various components. The pseudo-periods are estimated using the Fourier transform (FT) of what is not a pure sinusoidal curve. The duration of a cycle is on the order of the width of the analyzing time window (\eg \cite{claerbout1976}). As an example in Figure \ref{fig:09}, the FT of the lod component gives a value of 67.6 $\pm$ 24.4 yr compatible with the $\sim$90 yr Gleissberg cycle. The uncertainty is very large because in a 190 yr time window a 90 yr cycle can be observed only twice (Figure \ref{fig:09} top). But in the 600 yr time window from 1400 to 2000, the cycle occurs almost 7 times and the Fourier estimate of uncertainty is much improved (85.0 $\pm$ 8.5 yr for TRGRm in Figure \ref{fig:09} bottom). All the values for TRGRm listed in Table 01 are compatible with the lists for sunspots and Earth rotation in the sense of \Keogh \cite{keogh2001} and \Lin \cite{lin2003}.	
\newpage	
	

\section{The 11 yr Schwabe cycle  \label{sec:10}}
\begin{figure}[H]
		\centering{\includegraphics[width=0.8\columnwidth]{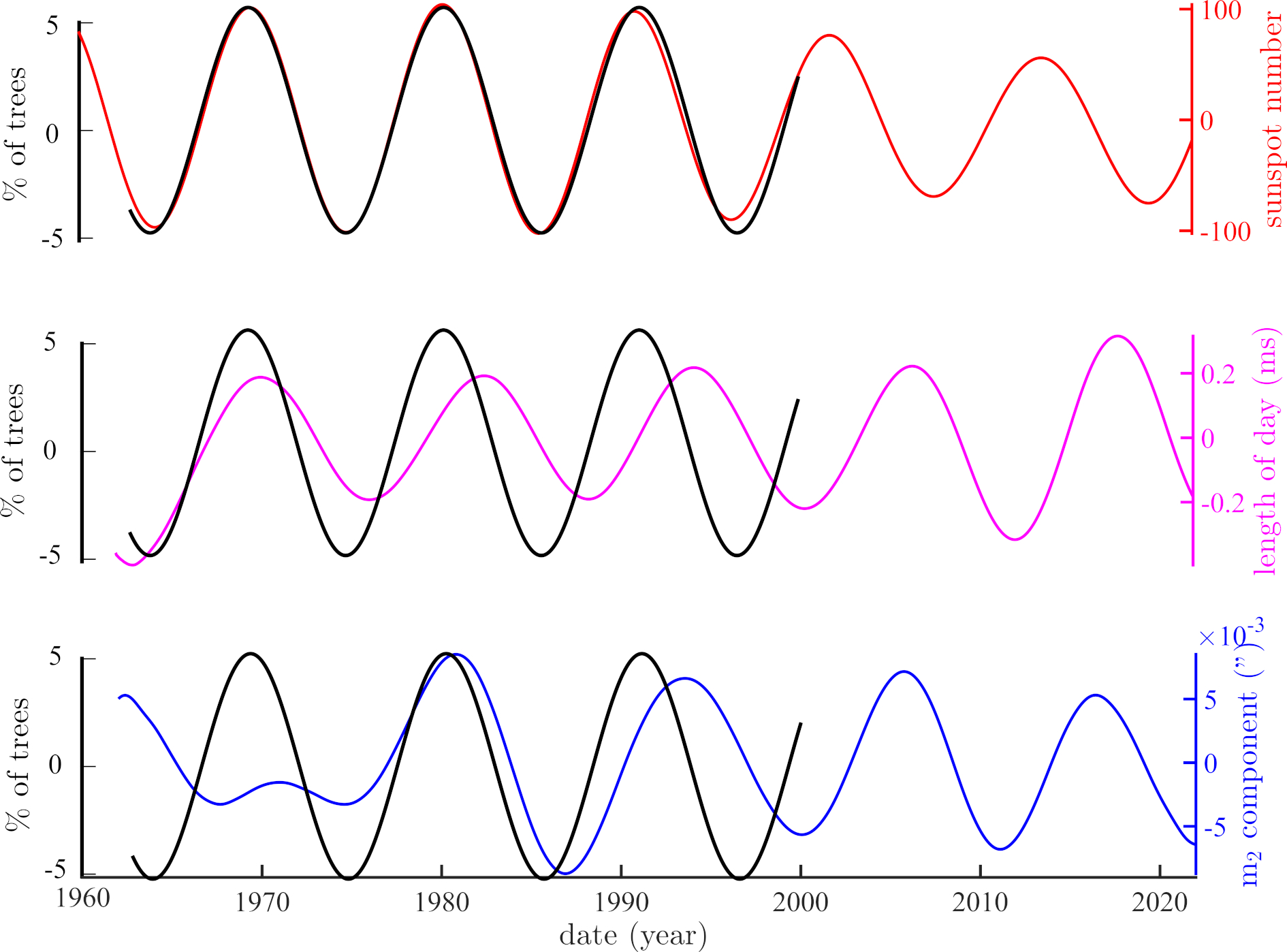}} 	
     	\caption{The black curve and left scale is the Schwabe cycle extracted by iSSA from the mean Tibetan tree ring series TRGRm in all three rows. The red curve in the top row is the Schwabe cycle component from sunspots, the red curve in the middle is the length of day and the blue curve at the bottom the $m_2$ pole motion component. All curves span the time of satellite measurements from 1962 to 2023.}
		\label{fig:12} 	
\end{figure}

		The \Schwabe \cite{schwabe1844} cycle is present in astronomy (\eg sunspots; \cite{usoskin2017,lemouel2020}) as well as in geophysics (\eg magnetic indices Dst and aa; \cite{lemouel2019b}).  We have also identified it in the length of day, $m_2$ component of pole motion and now in Tibetan Juniper growth rates TRGRm (this paper). Some of them are displayed in Figure 05. As already seen in Figure \ref{fig:12}, the 11-yr cycle recorded in the Junipers has been in phase with solar activity since the 1960s (Figure \ref{fig:12}, top). The phase agreement is less for length of day (with a drift in Figure \ref{fig:12}, middle) and the $m_2$  component (with a strong modulation in Figure \ref{fig:12}, bottom), whereas it was quite good with the three components displayed in Figures \ref{fig:09},\ref{fig:10} and \ref{fig:11}.
		
	We have already seen this sort of phase loss between the longer and shorter periods in the series of volcanic eruptions since 1700 \cite{lemouel2023b}, an intriguing question left to further work.	

\section{A long tree ring record and climate \label{sec:11}}
	As recalled in the introduction, the Dulan site data of \Fang \cite{fang2018} include a remarkable tree that covers almost entirely the time span from 357 to 2000AD, with only a small gap between 1050 and 1150AD. Having calculated the amplitudes, modulations and phases of the Schwabe and Gleissberg pseudo-cycles as determined from the tree ring series, one can re-scale the tree components as apparent sunspot number. The two re-scaled cycles are shown in Figure 06 from 357 to 2000AD.

	Importantly, the modulations of the \Schwabe and \Gleissberg "wavesforms" seen in the Dulan "ancestor" from 1400 to 2000 match those found for the median of all trees. So we can expect the single tree record to provide a valid estimate for the whole time range. This is true for all components listed in Appendix \ref{sec:A}. 
	
	We have superimposed in Figure \ref{fig:13} the Oort, Wolf, Spörer, Maunder, Dalton and Modern climate extrema. We find that each climate extremum corresponds to an extremum of the \Gleissberg cycle. The reciprocal statement does not hold. The \Gleissberg cycle appears to be strongly modulated with a period of $\sim$500-600 years for the recent segment between 1200 and 1800 (more or less the Little Ice Age), a previous segment between 700 and 1300, and a possible strong earlier one between $\sim$100-200 (?) and 700. Three 500-600 yr modulations of the Gleissberg cycle are apparent in Figure 06. The node near the small gap in the data is very close to the Medieval Climate Optimum. The Schwabe cycle is not as rich in internal structure. It is somewhat puzzling that it appears to have a much smaller effect on climate.

\begin{figure}[H]
		\centering{\includegraphics[width=0.8\columnwidth]{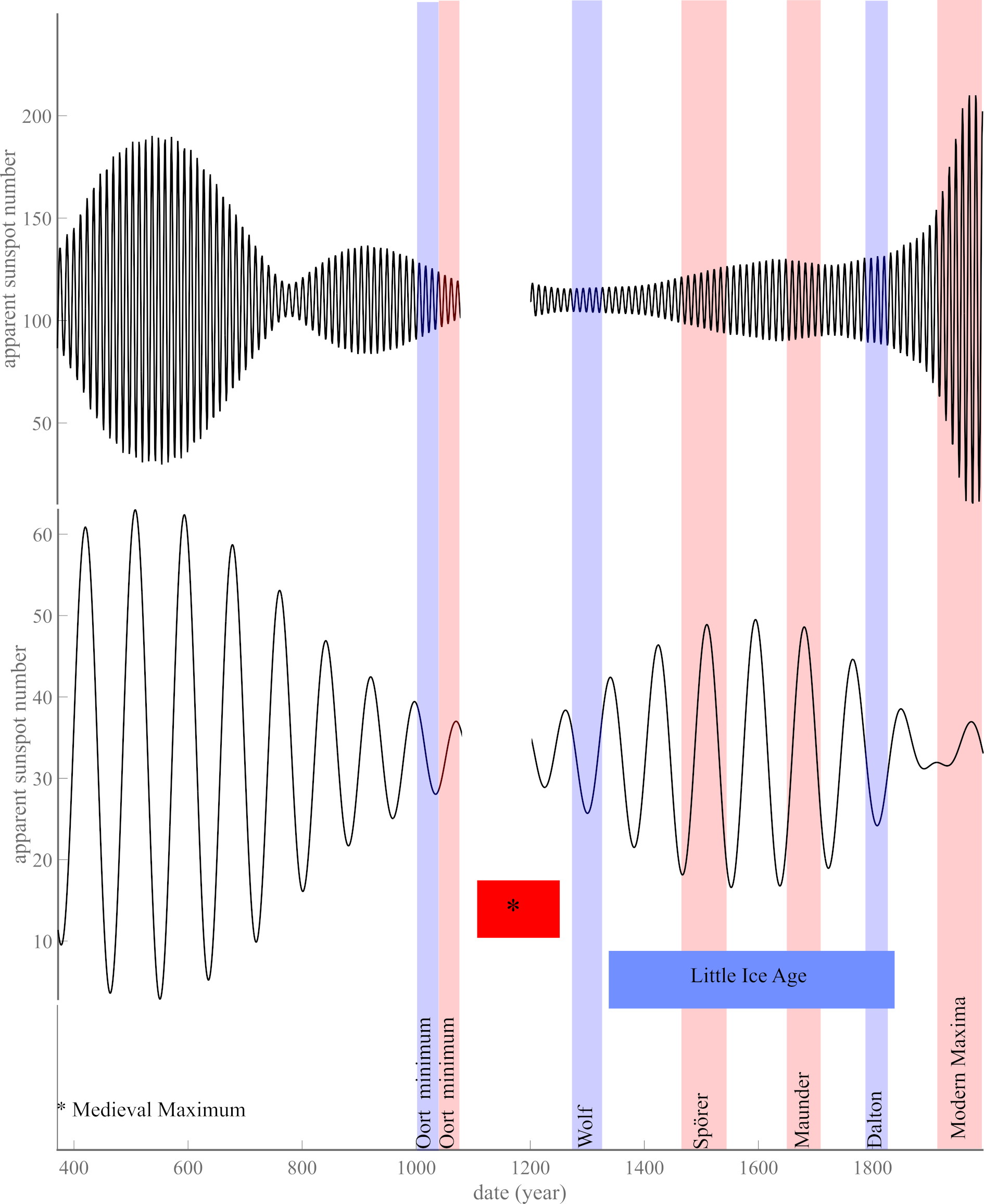}} 	
     	\caption{Comparison between apparent \Schwabe (top) and \Gleissberg (center) sunspot cycles and historical climates.}
		\label{fig:13} 	
\end{figure}

\section{Discussion and Conclusion\label{sec:12}}
	The results obtained from this analysis of the growth rates of long-lived Junipers from a remote area with little anthropic influence suggest the idea that such forests could be considered as key long-term astronomical, astrophysical and geophysical observatories. A wealth of information is recorded in the properties of thousands of tree-rings. 
\newline
		
	With the help of \Laplace \cite{laplace1799}'s work on mechanics, \Einstein \cite{einstein1912}’s physical explanation of photosynthesis, and \Milankovic \cite{milankovic1920}’s mathematical theory of climates, one can understand the physical link between the number of photons received at any position and time on Earth and the Earth’s motion through space. The link is captured by Milankovic’s equation \ref{eq:01}. The physical mechanics of \Laplace \cite{laplace1799} and \Lagrange \cite{lagrange1788} link variations of the inclination of the polar axis to its rotation velocity (\eg \cite{lopes2022a,lopes2022b,lopes2023a}) hence to the length of day. These are the same two parameters that are called obliquity and precession on much longer time scales (> 20 000 yr). In polar motion, the $\sim$11-yr term is very small compared to the main oscillations (\eg \cite{lopes2017,lopes2022b}). In contrast, the 90 yr Gleissberg, 22 yr and 30 yr terms are quite significant.
\newline
		
	Starting from the extensive data set of growth rates of tree rings from Tibetan Junipers by \Fang \cite{fang2018}, spanning from 57 to 2008 AD, we have built a median curve (TRGRm). The number of samples is stable in the six centuries between 1400 and 2000 so that the median estimates are not biased (Figures \ref{fig:01} and \ref{fig:02}) .  After having checked that the method of analysis we have selected, iSSA, yields consistent results that agree with the more "canonical" wavelet transform (Figures \ref{fig:05} and \ref{fig:06}), we have decomposed the TRGRm median curve into a sum of pseudo-cycles. The set of cycles spans from 3.3 yr up to more than 1000 years; taken together they capture 93\% of the total energy (variance) of the original TRGRm. Their values are listed in Table 01 and the components are shown in descending order in Appendix \ref{sec:A} (Figures \ref{fig:A01} to \ref{fig:A03}). An intriguing observation is that the decomposition of the median curve TRGRm is identical to that of global volcanism over the 3 centuries over which the two data sets are available (\cf \cite{lemouel2023a}). The phase variations of the two sets are comparable.
\newline
			
	We have compared the tree-ring cycles to the main cycles present in sunspots, that is the $\sim$11 yr \Schwabe \cite{schwabe1844} and $\sim$90-yr \Gleissberg \cite{gleissberg1939} cycles. Regarding the former one (Figure \ref{fig:07}), the components for tree rings and sunspots follow the same (growing) evolution since at least 1700. The phase history goes from perfect correlation between 1700 and 1740 to perfect phase opposition in the second half of the 19th century. In the recent decades, from 1970 to 2000, phase correlation returns. For the Gleissberg cycle (Figure \ref{fig:08}), the phase difference remains constant. When the tree ring curve is offset by 12 years, the superimposition is quasi-perfect, including the phase and amplitude modulations. This excellent coincidence of the solar and tree ring components explains in great part the success of dendro-chronology.
\newline
			
	We have next compared the tree-ring cycles with those in the length of day, available from 1832 onward. We have shown the results for the three longest pseudo-cycles, the Gleissberg $\sim$90 yr cycle (Figure \ref{fig:09}), $\sim$30 yr Markowitz oscillation (Figure \ref{fig:10}; \eg \cite{poma2000}) and the $\sim$22-yr Hale cycle (Figure \ref{fig:11}, \cite{raspopov2004}). The two series are in phase quadrature for the $\sim$90 and $\sim$33 yr cycles, and in phase opposition for the $\sim$22 yr cycle. 
\newline
		
	Since 1962 and the satellite era, there is little doubt that the tree-ring $\sim$11-yr \Schwabe cycle results from a causal chain that includes its solar and geophysical counterparts (Figure \ref{fig:12}). 
\newline	
	
	We have devoted a special section to analysis of the longest living tree in Dulan (357 to 2000 AD; Figure \ref{fig:13}). The main climate extrema and the well-known Medieval Climate Optimum, Little Ice Age and Modern Climate Optimum all seem to be mainly forced by variations in the envelope of the \Gleissberg cycle.
	
	Figure \ref{fig:13} contains useful information in terms of terrestrial climate. While studying the effect of glacier fluctuations in relation to the ages of Tibetan junipers, \Brauning \cite{brauning2006} demonstrated that both Tibet and Europe experienced a Little Ice Age (LIA) between the 14th and 19th centuries. The beginning of the Gleissberg cycle oscillation on Figure \ref{fig:13} aligns with the accepted starting date of the LIA in Europe around 1300-1303, as determined by \cite{pfister1998}. The LIA can actually be broken down into several phases, as noted by \Leroy \cite{leroy2010}, one of which is referred to as the "hyper-LIA." In Europe, this hyper-LIA (starting around 1560-1570) is characterized by "une forte poussée des glaciers alpins, du fait d’un régime assez fréquent d’étés dépressionnaires et pourris, frais, défavorables à l’ablation des glaces, mais aussi d’hivers froids et probablement fort neigeux. Cette poussée glaciaire culmine une première fois pendant les années 1590, au cours desquelles les glaciers de la région de Chamonix et Grindelwald culbutent des chapelles et des hameaux situés en position quasi frontale et marginale"\footnote{$\ldots$ a strong episode of alpine glacier "push", due to a rather frequent regime of rotten, cool Summer depressions, unfavorable to ice removal, but also to cold and likely very snowy Winters. This glacial push had a first maximum during the 1590s, when glaciers in the regions of Chamonix and Grindelvald overturned chapels and hamlets located almost at their front and on the margins.}  (\cf. \cite{leroy2010}, page 20). Indeed, we observe that the three most significant oscillations of the Gleissberg cycle occur between approximately 1560-1570 and 1640-1645, with a peak around 1590. Referring to Europe, as mentioned in \Leroy \cite{leroy2010}, page 21: "À partir de 1644-1645, les glaciers chamoniards, et surtout le glacier inférieur de Grindelwald, amorcent un léger retrait, qui reste cependant, lui aussi, dans les limites très étoffées du PAG."\footnote{Beginning in 1644-1645, glaciers around Chamonix and mainly the lower Grindelwald glacier engaged in a slight retreat that itself remains in the broad limits of the LIA.} Thus, the Gleissberg cycle extracted from the oldest individual Juniper in Dulan, matches the observations reported by \Leroy.
\newline
		
	In Europe, before the LIA, there was a period known as the Medieval Warm Period (MWP) between the 9th and 13th centuries  (\cf \cite{leroy2010}, page 24). These dates align perfectly (despite the data gap) with the evolution of the envelope of our Gleissberg cycle. During the same period, in Tibet, the amplitude modulations of this $\sim$90-year cycle appear to reach their lowest values. Whether this warm period was a global phenomenon is still debated. Observations in different parts of Earth (\eg \cite{villalba1990,cook1991,graumlich1993,hughes1994}) are in favor of a global extension.
\newline
			
	One should underline the fact the Schwabe cycle appears to have little effect on climate. Indeed, the amplitude of the 11-year cycle appears to increase gradually since the Medieval Warm Period without ever showing any correlation with the different climatic cycles discussed in the previous paragraph.
\newline
			
	The \Milankovic \cite{milankovic1920} mathematical theory of climate (and equation 01) allows one to relate climate change and length of day, through changes in inclination of Earth’s rotation axis and solar insolation. The set of pseudo-periods that are evidenced in the Tibetan tree ring growth rates simply corresponds to short period \Milankovic \cite{milankovic1920} cycles. Their sequence forms what we have called a specific spectral signature SSS. Although we do keep in mind that correlation may not imply causality, we have shown that recognition of the SSS likely tells us that geophysical series on Earth have recorded changes driven by changes in the rotation axis, themselves driven by exchanges of angular momentum involving the Sun and Jovian planets. In that case the reality and sense of causality is unmistakable.

\appendix
\counterwithin{figure}{section}
\section{Pseudo-cycles extracted from TRGRm \label{sec:A}}    %% Appendix A
\begin{figure}[H]
		\centering{\includegraphics[width=0.8\columnwidth]{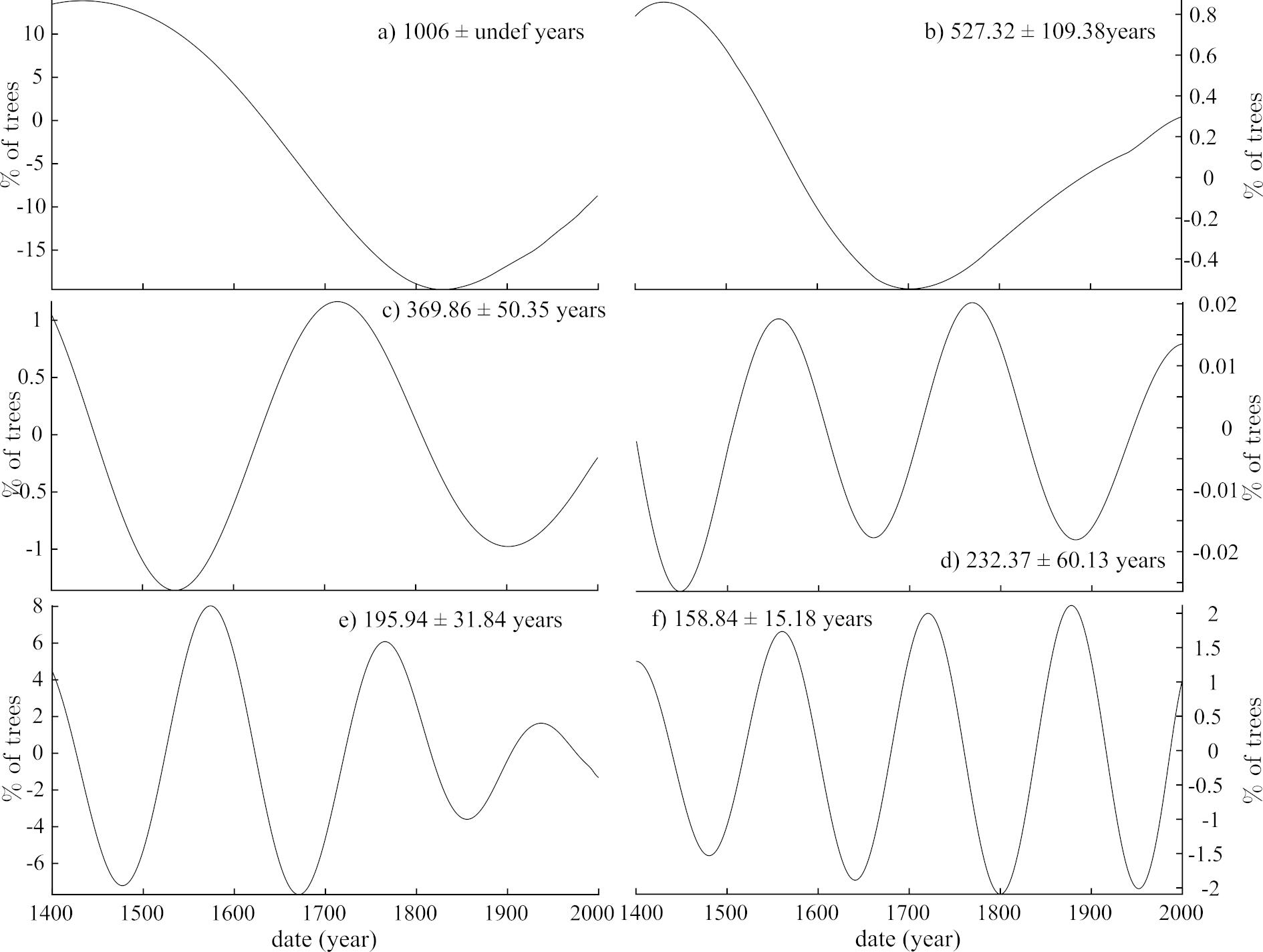}} 	
     	\caption{Pseudo-cycles with periods (in years): $\sim$1006, $\sim$537, $\sim$369, $\sim$232, $\sim$195 and $\sim$158 extracted from the series of Tibetan Juniper tree-ring growth rates (1400 to 2000 AD).}
		\label{fig:A01} 	
\end{figure}	
\begin{figure}[H]
		\centering{\includegraphics[width=0.8\columnwidth]{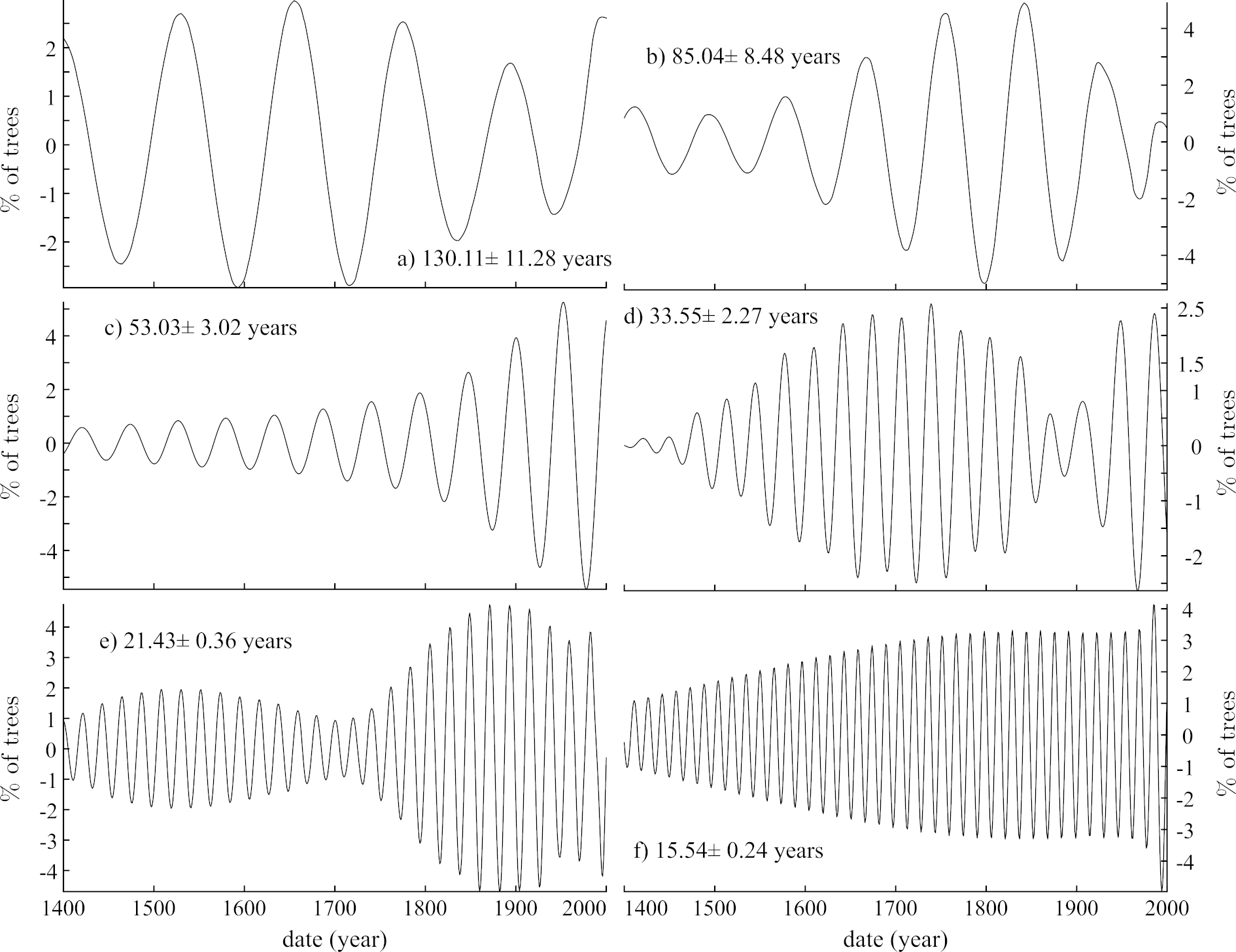}} 	
     	\caption{seudo-cycles with periods (in years): $\sim$130, $\sim$85, $\sim$53, $\sim$33, $\sim$21 and $\sim$15 extracted from the series of Tibetan Juniper tree-ring growth rates (1400 to 2000 AD).}
		\label{fig:A02} 	
\end{figure}	
\begin{figure}[H]
		\centering{\includegraphics[width=0.8\columnwidth]{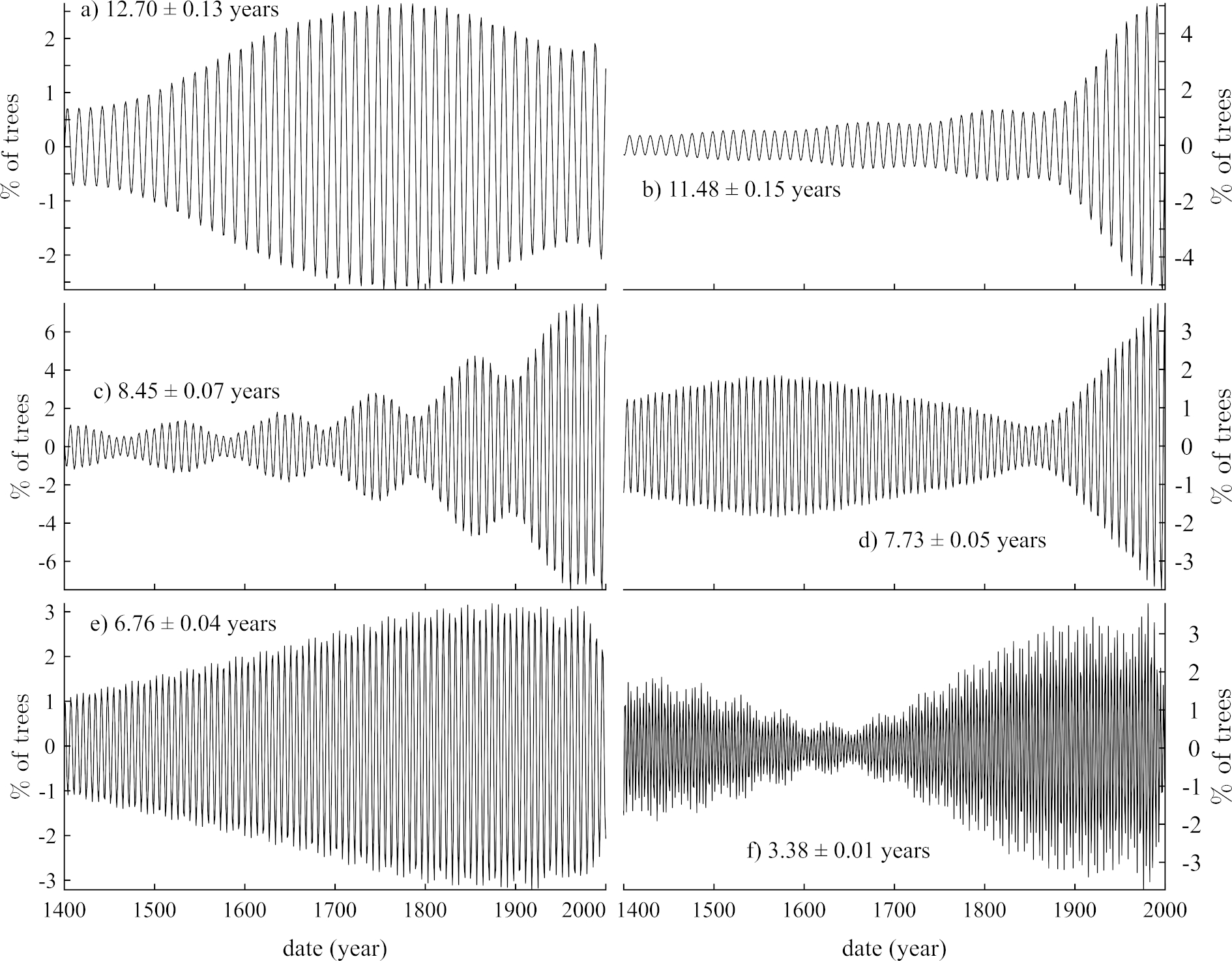}} 	
     	\caption{Pseudo-cycles with periods (in years): $\sim$12, $\sim$11, $\sim$9, $\sim$8, and $\sim$3.3 extracted from the series of Tibetan Juniper tree-ring growth rates (1400 to 2000 AD).}
		\label{fig:A03} 	
\end{figure}	

\newpage
\bibliographystyle{ieeetr}
\bibliography{tibet.bib}

\end{document}